\documentclass[prb,twocolumn,floatfix,showpacs,citeautoscript]{revtex4}

\usepackage{amsmath}
\usepackage{amssymb}
\usepackage{amsfonts}
\usepackage{graphicx,graphics}
\usepackage{bm}  

\newcommand{\ket}[1]{\lvert #1\rangle}
\newcommand{\bra}[1]{\langle#1 \rvert}
\newcommand{\abs}[1]{\lvert #1 \rvert}
\newcommand{\expect}[1]{\langle #1\rangle}
\newcommand{\braket}[2]{\langle #1 \rvert #2\rangle}

\newcommand{\bk}{\mathbf{k}}
\newcommand{\bq}{\mathbf{q}}

\begin{document}

\title{Unraveling the acoustic electron-phonon interaction in graphene}

\author{Kristen Kaasbjerg}
\email{cosby@fys.ku.dk}
\author{Kristian S. Thygesen}
\author{Karsten W. Jacobsen}
\affiliation{Center for Atomic-scale Materials Design (CAMD),\\
             Department of Physics, Technical University of Denmark}
\date{\today}

\begin{abstract}
  Using a first-principles approach we calculate the acoustic electron-phonon
  couplings in graphene for the transverse (TA) and longitudinal (LA) acoustic
  phonons. Analytic forms of the coupling matrix elements valid in the
  long-wavelength limit are found to give an almost quantitative description of
  the first-principles based matrix elements even at shorter wavelengths. Using
  the analytic forms of the coupling matrix elements, we study the acoustic
  phonon-limited carrier mobility for temperatures $0-200$~K and high carrier
  densities of $10^{12}-10^{13}$~cm$^{-2}$. We find that the intrinsic
  \emph{effective} acoustic deformation potential of graphene is $\Xi_\text{eff}
  = 6.8$~eV and that the temperature dependence of the mobility $\mu \sim
  T^{-\alpha}$ increases beyond an $\alpha = 4$ dependence even in the absence
  of screening when the full coupling matrix elements are considered. The large
  disagreement between our calculated deformation potential and those extracted
  from experimental measurements ($18-29$~eV) indicates that additional or
  modified acoustic phonon-scattering mechanisms are at play in experimental
  situations.
\end{abstract}


\maketitle

\section{Introduction}

Since the experimental realization of graphene~\cite{Firsov:Science}, its
electronic properties and their understanding have been studied extensively both
experimentally and theoretically~\cite{Geim:Graphene,RMP:Graphene,Sarma:RMP}.
While the intrinsic carrier mobility of graphene is predicted to be
exceptionally high, the experimental reality in substrate supported graphene
involving charged impurities, electron-hole puddles, surface-optical phonons of
the substrate, and disorder typically results in strongly reduced mobilities
compared to the expected intrinsic
value~\cite{Fuhrer:GrapheneOnSiO2,Sarma:ScreeningInduced}. Together with
scattering on acoustic phonons which manifests itself in a linear temperature
dependence of the mobility at higher temperatures, these extrinsic scattering
mechanisms typically dominate the mobility. The linear temperature dependence
characteristic for acoustic phonon scattering has so far been observed in both
supported~\cite{Fuhrer:GrapheneOnSiO2,Zhu:Deposition,Hone:BNSubstrate,Kim:Controlling,Ozyilmaz:ElectrolyteGate}
and suspended~\cite{Kim:Suspended} graphene samples.

With the recent improvements in sample fabrication, the relative role of
acoustic phonon scattering must be expected to become increasingly important in
future devices. For example, samples with the commonly used SiO$_2$ substrate
replaced by hexagonal boron nitride (h-BN) which has a lattice constant very
close to that of graphene and an almost atomically flat surface with strongly
reduced disorder~\cite{LeRoy:STMhBN,Crommie:LocalBN}, have shown highly improved
transport characteristics with mobilities approaching that of suspended
graphene~\cite{Hone:BNSubstrate,Shepard:HallEffect,Wees:Transfer}. Furthermore,
the high energy of the surface-optical phonons of h-BN results in a significant
reduction of surface-optical phonon
scattering~\cite{Guinea:SubstrateLimited,Jena:High-k,Kim:SurfacePolar} that for
commonly used gate oxides starts to dominate the mobility around $T \sim
150-200$~K~\cite{Fuhrer:GrapheneOnSiO2,Zhu:Deposition}.

When the mobility is dominated by acoustic phonon scattering, two transport
regimes separated by the Bloch-Gr{\"u}neisen (BG) temperature $T_\text{BG}= 2
\hbar k_F c_\text{ph} / k_\text{B}$ can be identified~\cite{Ziman}. Here, $k_F$
is the Fermi wave vector, $c_\text{ph}$ the sound velocity and $k_\text{B}$ the
Boltzmann constant ($T_\text{BG} \sim 57$~K $\sqrt{n}$ for the LA phonon with
$n$ measured in units of $10^{12}$~cm$^{-2}$). Since the BG temperature
corresponds to the acoustic phonon energy $\hbar\omega_\bq=\hbar c_\text{ph} q$
for full backscattering at the Fermi level, short wavelength acoustic phonons
are frozen out at temperatures $T < T_\text{BG}$ restricting scattering
processes to small scattering angles. The restricted phase space available for
phonon scattering at $T < T_\text{BG}$ results in a transition from the linear
$\rho \sim T$ behavior of the resistivity in the high-temperature regime to a
stronger $\rho \sim T^\alpha$ temperature dependence in the BG regime where
$\alpha = 4$ ($\alpha = 6$) in the absence (presence) of screening by the
carriers themselves~\cite{Sarma:Acoustic,Sarma:Chirality}. The BG behavior in
the temperature dependence of the mobility (resistivity) has recently been
observed experimentally~\cite{Kim:Controlling}.

Existing theoretical~\cite{Sarma:Acoustic,Sarma:Chirality} and
experimental~\cite{Fuhrer:GrapheneOnSiO2,Kim:Suspended,Zhu:Deposition,Hone:BNSubstrate,Kim:Controlling}
studies of acoustic phonon-limited transport in graphene often parametrize the
interaction with acoustic phonons in terms of a coupling to a single
\emph{effective} acoustic phonon. The associated deformation potential coupling
constant extracted from the experimentally measured temperature dependence of
the resistivity range from $\sim 18 -
29$~eV~\cite{Fuhrer:GrapheneOnSiO2,Kim:Suspended,Zhu:Deposition,Hone:BNSubstrate,Kim:Controlling}.
On the other hand, theoretical studies of the acoustic electron-phonon coupling
yield much lower values on the order of
$3-4.5$~eV~\cite{Kim:ElPhGraphene,Avouris:Inelastic}. At the same time,
different forms of the coupling matrix element are used in theoretical
studies~\cite{Sarma:Acoustic,Avouris:Inelastic} making a direct comparison of
the different values of the deformation potential difficult.

Even though the effect of acoustic phonon scattering on the carrier mobility in
graphene has been studied widely in the
literature~\cite{Mahan:ElphGraphene,Guinea:ElectronicTransport,Sarma:Acoustic,Guinea:Flexural,Ferry:Intrinsic,Kim:ElPhGraphene},
a complete study considering the full details of the coupling matrix element is
still lacking. The purpose of the present study is to provide a detailed
analysis of the acoustic electron-phonon interaction in graphene and to
establish the intrinsic value of the effective deformation potential. We shall
focus on supported graphene where the flexural phonons are quenched and hence
include only the transverse (TA) and longitudinal (LA) acoustic phonons. We use
a first-principles method to calculate the electron-phonon
interaction~\cite{Kaasbjerg:MoS2,GPAW} supported by the group-theoretical
considerations of Ref.~\onlinecite{Manes:Symmetry}. We then study the intrinsic
phonon-limited mobility using a Boltzmann equation approach in the temperature
regime $0-200$~K and for high carrier densities $n \sim
10^{12}-10^{13}$~cm$^{-2}$ where screening by the carriers suppresses other
scattering mechanisms. Using the full coupling matrix elements, we find that a
temperature dependence with $\alpha > 4$ occurs even in the absence of
screening.

\section{Theory}

In the following, the carriers in graphene are described by massless Dirac
fermions with linear dispersion $\varepsilon_\bk = \hbar v_F k$ where $v_F \sim
1.0\times10^6$~m/s is the Fermi velocity. Within the Boltzmann equation
approach~\cite{Guinea:ElectronicTransport,Sarma:CarrierTransport,Sarma:ScreeningInduced},
the mobility in graphene in the presence of (quasi) elastic scattering
mechanisms is given by
\begin{equation}
  \label{eq:mobility}
  \mu_{xx} = \frac{\sigma_{xx}}{ne} 
           = \frac{e v_F^2 \expect{\tau_\bk}}{2}
\end{equation}
where $\sigma_{xx}$ is the conductivity, $n$ is the two-dimensional carrier
density and the density-of-states averaged relaxation time (in units of time per
energy) is defined by 
\begin{equation}
  \label{eq:tau_averaged}
  \expect{\tau_\bk} = \frac{1}{n} \int \! d\varepsilon_\bk \; \rho(\varepsilon_\bk) 
          \left(- \frac{\partial f}{\partial \varepsilon_\bk} \right)
          \tau_\bk .
\end{equation}
Here, $\rho(\varepsilon_\bk) = (g_s g_v/2\pi\hbar^2) \varepsilon_\bk / v_F^2$ is
the density of states of the graphene layer and $g_s=2$ and $g_v=2$ are the spin
and valley degeneracies, respectively. At low temperatures and high carrier
densities where $\varepsilon_F \gg k_\text{B} T$ this yields $\mu_{xx} \approx e
v_F^2 \tau_{_{k_F}} / \varepsilon_F$.

In the case of acoustic phonon scattering which can be treated as a quasielastic
process, the relaxation time for each of the acoustic phonons (TA and LA) is
given by~\cite{Sarma:Acoustic}
\begin{equation}
  \label{eq:tau_acoustic}
  \frac{1}{\tau_{\bk\lambda}} = \sum_{\bk'} 
      \left( 1 - \cos{\theta_{\bk,\bk'}} \right) P_{\bk\bk'}^\lambda 
       \frac{1 - f_{\bk'}}{1 - f_\bk}   ,
\end{equation}
where $\theta_{\bk,\bk'}$ is the scattering angle and $f_\bk=f(\varepsilon_\bk)$
the Fermi function. The transition matrix element is given by
\begin{align}
  \label{eq:P_acoustic}
  P_{\bk\bk'}^\lambda = \frac{2\pi}{\hbar} \sum_\bq
    & \left\vert g_{\bq\lambda} \right\vert^2 
    \bigg[ 
    N_{\bq\lambda} 
    \delta(\varepsilon_{\bk'} - \varepsilon_\bk - \hbar\omega_{\bq\lambda}) \bigg. \nonumber \\
    & \quad + \bigg. \left(1 + N_{\bq\lambda} \right) 
     \delta(\varepsilon_{\bk'} - \varepsilon_{\bk} + \hbar\omega_{\bq\lambda})
      \bigg]  .
\end{align}
Here, $g_{\bk\bq}^\lambda$ is the electron-phonon coupling, $\lambda$ denotes
the acoustic phonon branches, and $\hbar\omega_{\bq\lambda}$ the acoustic phonon
energy. In the following, these quantities are calculated from first-principles.
The phonons are assumed to be in equilibrium and populated according to the
Bose-Einstein distribution function
$N_{\bq\lambda}=N(\hbar\omega_{\bq\lambda})$. As scattering on both the TA and
LA phonon is considered here, the total relaxation time for the $K,K'$-valleys
is given by the sum of the individual phonon contributions as $\tau^{-1} =
\sum_\lambda \tau_{\lambda}^{-1}$. As we show in the following section, the
matrix elements of the electron-phonon coupling differ in the $K$ and $K'$
valleys (see e.g. Fig.~\ref{fig:couplings}). The Boltzmann equation must
therefore be solved explicitly in both valleys. In the absence of intervalley
scattering which couples the distribution functions in the two valleys, this can
be done by considering the two valleys separately. In this case, the relaxation
time entering the expression for the mobility in Eq.~\eqref{eq:mobility} becomes
the valley-averaged relaxation time $\tau = (\tau_{_K} + \tau_{_{K'}})/2$, where
$\tau_{_{K/K'}}$ is the total relaxation time in the individual valleys.
Screening of the electron-phonon interaction by the carriers
themselves~\cite{Guinea:Dynamical,Sarma:Dielectric} has been considered
elsewhere~\cite{Sarma:Chirality} and will here be neglected.

While analytic considerations have been given in
Refs.~\onlinecite{Sarma:Acoustic,Sarma:Chirality}, we will in the present work
resort to a numerical evaluation~\cite{footnote1,footnote2} of the relaxation
time in Eqs.~\eqref{eq:tau_acoustic} and~\eqref{eq:P_acoustic}. This allows us
to study the acoustic phonon-limited mobility in graphene with the full coupling
matrix elements which have a more complex angular dependence than most often
assumed (see e.g. Ref.~\onlinecite{Sarma:Acoustic}). The numerical approach also
allows for a unified treatment of the high-temperature ($T > T_\text{BG}$) and
Bloch-Gr{\"u}neisen ($T < T_\text{BG}$) regimes. We note, however, that in the
low-temperature regime where $\hbar\omega_\bq \sim k_\text{B}T$, it is crucial
that the phonon energy is retained in the Fermi function in
Eq.~\eqref{eq:tau_acoustic}. In the high-temperature regime this requirement can
be relaxed and the relaxation time can be put on a simple analytic form.

\section{Interaction with acoustic phonons}

In the following we use a first-principles DFT approach based on a fully
microscopic description of both the electronic Bloch states and the phonons to
calculate the acoustic electron-phonon couplings in
graphene~\cite{Kaasbjerg:MoS2,GPAW,footnote3}. The calculated values for the
sound velocities $c_\lambda$ for the TA and LA phonons are reported in
Table~\ref{tab:parameters} together with other parameters used in this work. Due
to their high phonon energies ($ > 100$~meV), acoustic intervalley and optical
phonons do not play a role in the considered temperature range and can therefore
be neglected.

The interaction between charge carriers and the acoustic phonons in graphene can
be written in the general form
\begin{equation}
  \label{eq:g}
  g_{\bk\bq}^\lambda = \sqrt{\frac{\hbar}{2A\rho\omega_{\bq\lambda}}}
                       M_{\bk\bq}^\lambda ,
\end{equation}
where $A$ is the area of the graphene layer, $\rho$ is the mass density and
$M_{\bk\bq}^\lambda = \bra{\bk+\bq} \delta V_{\bq\lambda} \ket{\bk}$ is the
coupling matrix element for scattering between the two Bloch states $\bk$ and
$\bk+\bq$ due to a phonon with wave vector $\bq$, branch index $\lambda$ and
frequency $\omega_{\bq\lambda}=c_\lambda q$ where $c_\lambda$ is the sound
velocity for the acoustic branches. The coupling is mediated by the change
$\delta V_{\bq\lambda}$ in the microscopic crystal potential due to a unit
displacement of the atoms along the mass-scaled normal mode vector
$\mathbf{e}_{\bq\lambda}$. Due to the full microscopic treatment of both
electrons and phonons, Umklapp processes involving reciprocal lattice vectors
are included in the coupling matrix element $M_{\bk\bq}^\lambda$.
\begin{table}[!b]
\begin{ruledtabular}
\begin{tabular}{lcc}
Parameter &  Symbol  & Value  \\ 
\hline
Lattice constant               &   $a$                  &   2.46~{\AA} (LDA)          \\
Ion mass density               &   $\rho$               &   $7.6\times 10^{-8}$~g/cm$^2$ \\
Fermi velocity                 &   $v_F$                &   $1.0\times 10^6$~m/s \\
Transverse sound velocity      &   $c_\text{TA}$        &   $14.1 \times 10^3$~m/s  \\
Longitudinal sound velocity    &   $c_\text{LA}$        &   $21.2 \times 10^3$~m/s  \\
Electron-phonon couplings      &                        &                  \\
Transverse                     &   $\beta_\text{TA}$    &   $2.8$~eV  \\
Longitudinal                   &   $\alpha_\text{LA}$   &   $2.8$~eV  \\
Longitudinal                   &   $\beta_\text{LA}$    &   $2.5$~eV  \\
Effective coupling parameters  &                        &                  \\
Sound velocity                 &   $c_\text{eff}$       &   $20.0 \times 10^3$~m/s  \\
Deformation potential          &   $\Xi_\text{eff}$     &   $6.8$~eV  \\
\end{tabular}
\end{ruledtabular}
\caption{Material parameters for graphene used in the present work. The phonon
  related parameters have been obtained from first-principles as described in
  the text. The calculated sound velocities are in excellent agreement with the
  values reported in Ref.~\onlinecite{Guinea:Flexural}.}
\label{tab:parameters}
\end{table}

\begin{figure}[!t]
  \begin{minipage}{1.0\linewidth}
    \includegraphics[width=0.73\linewidth]{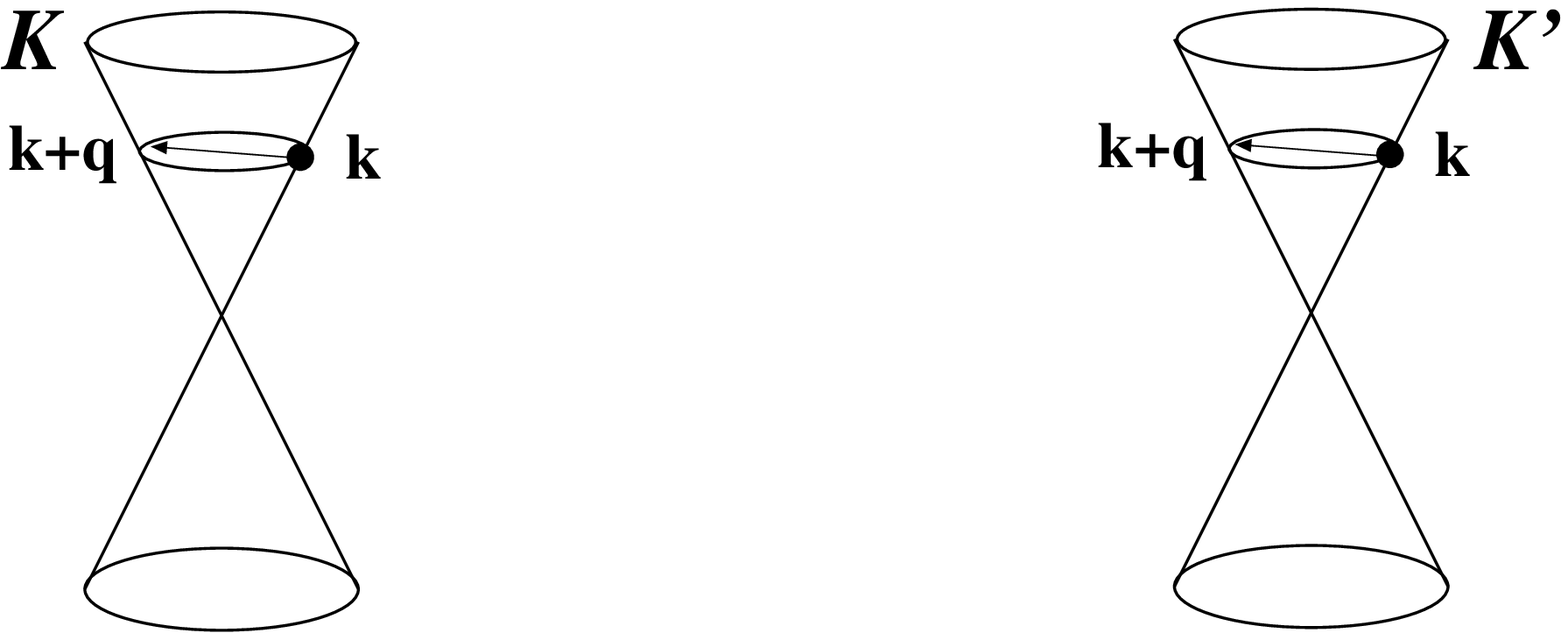}
  \vspace{.3cm}
  \end{minipage}
  \begin{minipage}{1.0\linewidth}
    \includegraphics[width=0.49\linewidth]{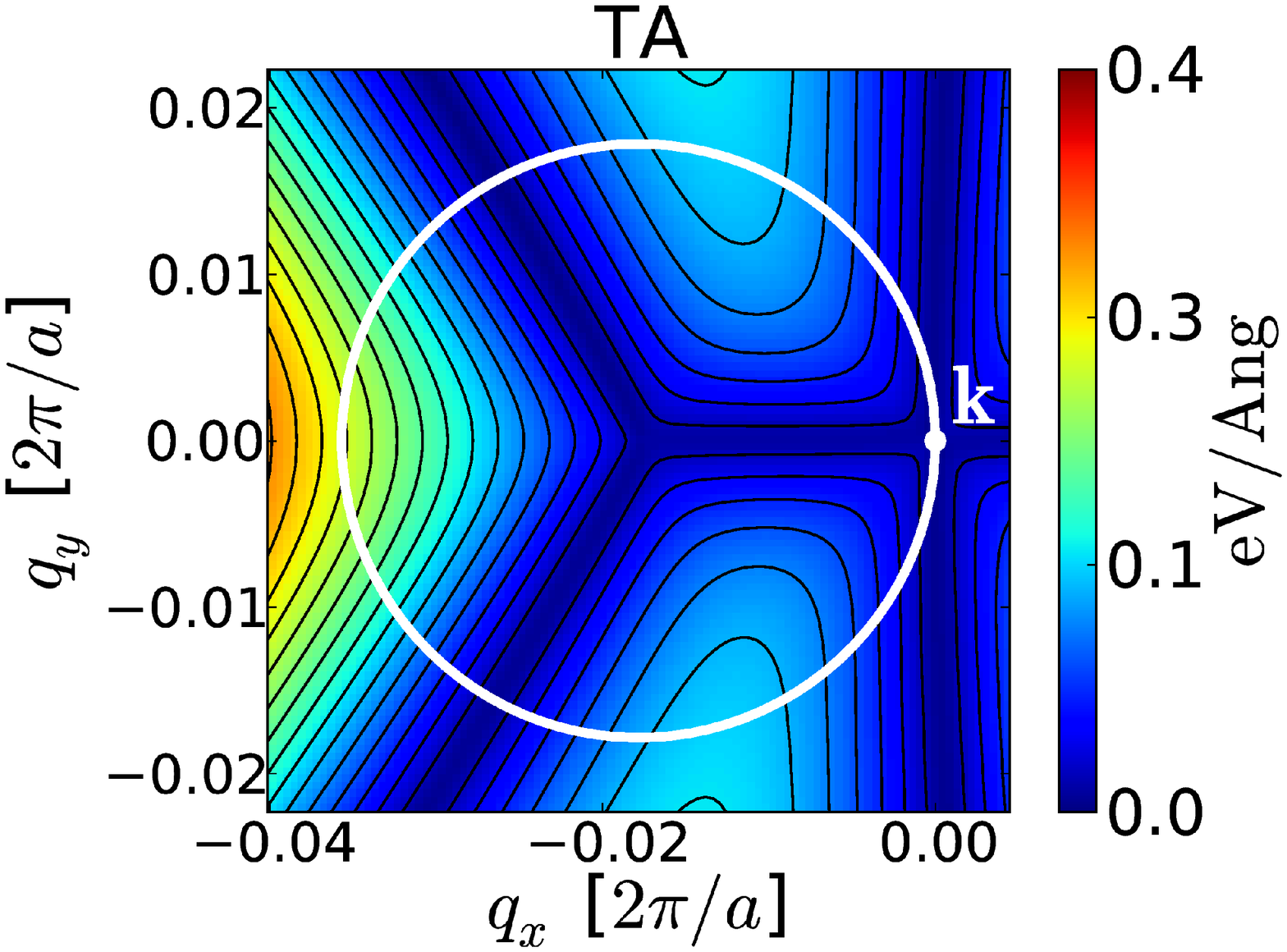}
    \includegraphics[width=0.49\linewidth]{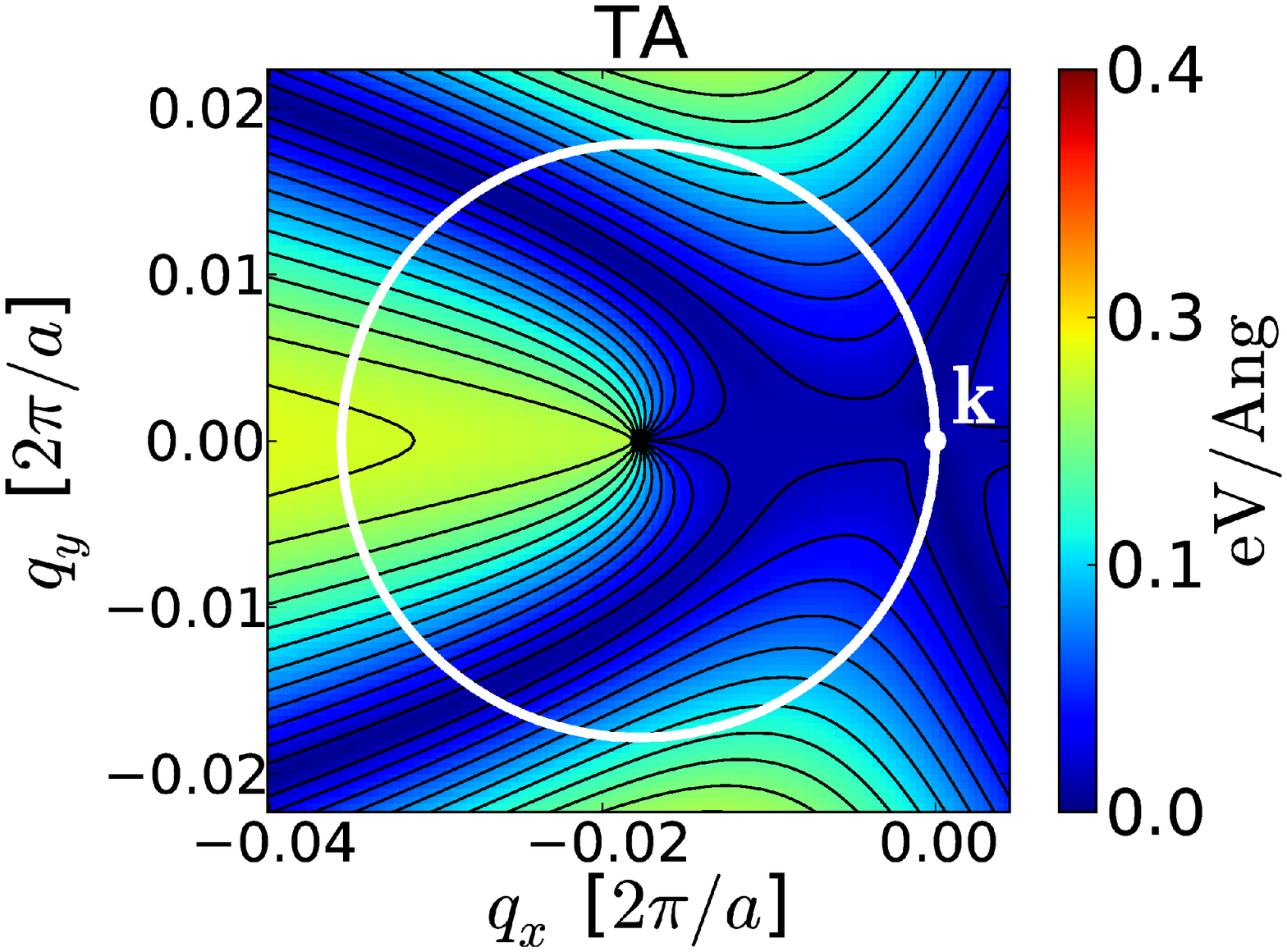}
  \end{minipage}
  \begin{minipage}{1.0\linewidth}
    \includegraphics[width=0.49\linewidth]{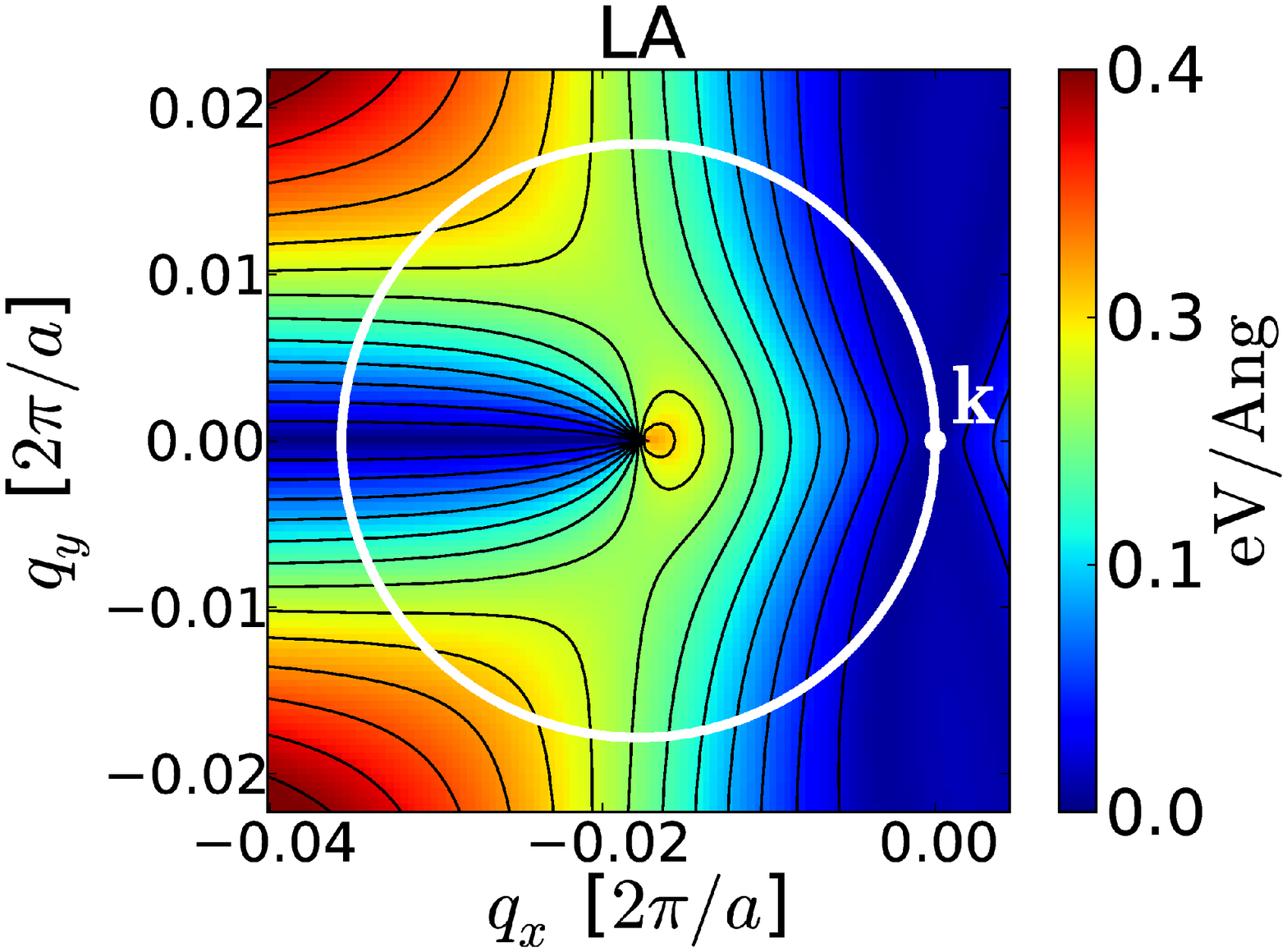}
    \includegraphics[width=0.49\linewidth]{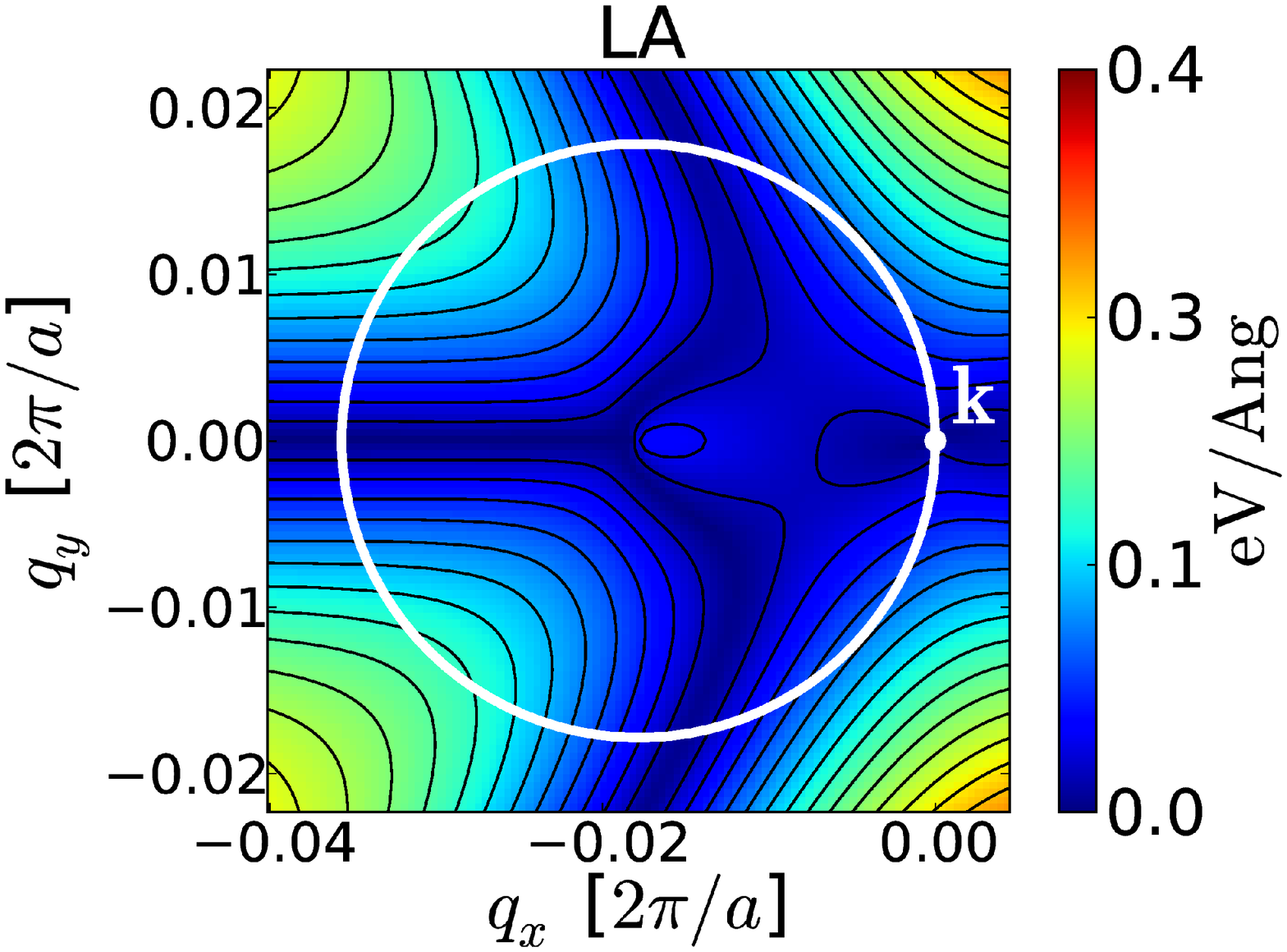}
  \end{minipage}
  \caption{(Color online) Electron-phonon couplings to the acoustic TA and LA
    phonons in the $K,K'$-valleys (left and right columns, respectively) of
    graphene. The contour plots show the absolute value of the coupling matrix
    elements $\abs{M_{\bk\bq}^\lambda}$ for a carrier energy of 300~meV as a
    function of the two-dimensional phonon wave vector $\bq$. The (white)
    circles correspond to $\bk + \bq$ vectors lying on the constant energy
    surfaces of the Dirac cones given by the energy $\varepsilon_\bk$ of the
    initial state in $\bk$ as sketched in the top row.}
\label{fig:couplings}
\end{figure}

Figure~\ref{fig:couplings} shows the absolute value of the coupling matrix
elements $M_{\bk\bq}^\lambda$ to the TA and LA phonons in the $K,K'$ valleys as
a function of the two-dimensional phonon wave vector $\bq$. The matrix elements
in the two valleys are related through time-reversal symmetry as
${M_{\bk\bq}^{K\lambda}}^* = M_{-\bk,-\bq}^{K'\lambda}$. This implies that
carriers in the two valleys traveling in the same direction
experience different electron-phonon couplings. It should be emphasized that the
calculated matrix elements include the full symmetry of both the electronic
Bloch states and the phonon modes as given by the complete microscopic
description.

In order to emphasize the effect of the chirality of the carriers in graphene,
the initial carrier state $\bk$ is located on the right side of the Dirac cones
300~meV above the $K,K'$-points as indicated by the sketch in the top of the
figure. As is evident from the figure, both the TA and LA phonons couple to the
carriers with similar coupling strengths. While backscattering is suppressed for
the LA mode, the situation is reversed for the TA mode where forward scattering
is suppressed. In addition to suppression of forward and backscattering, other
directions with complete suppression of scattering also appears. This is a
consequence of the inclusion of the symmetry of both phonons and electronic
states.

In the following, the first-principles coupling matrix elements are analyzed
using the group-theoretical analysis of the electron-phonon interaction
presented in Ref.~\onlinecite{Manes:Symmetry}. In the long-wavelength limit, the
TA and LA phonons are strictly transverse and longitudinal,
respectively~\cite{footnote4}, and the electron-phonon interaction has a simple
analytic representation in the two-dimensional pseudospin
basis~\cite{Manes:Symmetry}. Using the results of
Ref.~\onlinecite{Manes:Symmetry}, the coupling matrix elements can expressed in
terms of the angles $\theta_\bk$, $\theta_\bq$ and $\theta_{\bk+\bq}$ of the
involved wave vectors. Including contributions of order $O(q)$, we find that the
coupling matrix elements in the long-wavelength limit takes the following form
in the $K$-valley,
\begin{align}
  \label{eq:M_TA_abinitio}
  \abs{M_{\bk\bq}^\text{TA}} = q \beta \times 
  \left\vert \sin\left( 2 \theta_\bq + 
      \frac{\theta_\bk + \theta_{\bk + \bq}}{2} \right) \right\vert 
\end{align}
and
\begin{align}
  \label{eq:M_LA_abinitio}
  \abs{M_{\bk\bq}^\text{LA}} & = q \times \left\vert \alpha
      \cos\left( \frac{\theta_{\bk + \bq} - \theta_\bk}{2} \right) \right. \nonumber \\ 
      & \quad\quad\quad \left. + \beta 
      \cos\left( 2 \theta_\bq + 
        \frac{\theta_\bk + \theta_{\bk + \bq}}{2} \right) \right\vert ,
\end{align}
for the TA and LA phonons, respectively. For the LA phonon, the first and second
terms originate from the deformation potential and the \emph{gauge field}
coupling mechanisms, respectively. The TA phonon couples only through the
latter~\cite{Manes:Symmetry,Oppen:Flexural,Oppen:Suspended}. Since the
interaction is Coulombic in nature, the overall couplings given in
Eqs.~\eqref{eq:M_TA_abinitio} and~\eqref{eq:M_LA_abinitio} are here referred to
as deformation potential couplings.

With the coupling parameters listed in Table~\ref{tab:parameters}, we find that
the analytic expressions for the electron-phonon interaction in
Eqs.~\eqref{eq:M_TA_abinitio} and~\eqref{eq:M_LA_abinitio} to a high degree
reproduce the first-principles matrix elements for electron energies up to $\sim
750$~meV. As the analytic coupling matrix elements are based on the phonon modes
in the long-wavelength limit, the agreement is slightly worsened at shorter
wavelengths where the mode vectors deviate from the long-wavelength
modes~\cite{footnote4}. This is most pronounced for the TA phonon. As the LA
phonon retains its long-wavelength character far out in the Brillouin zone, the
agreement between the coupling matrix elements here remains quantitative even at
shorter wavelengths. In the BG regime where short wavelength phonons are frozen
out, we note that it is the long-wavelength limit of the coupling matrix
elements that governs the scattering of carriers.

Often scattering on acoustic phonons is described by coupling to a single
\emph{effective} phonon mode with a coupling matrix element given
by~\cite{Guinea:ElectronicTransport,Sarma:Acoustic}
\begin{equation}
  \label{eq:M_Sarma}
  M_{\bk\bq}^\text{eff}
  = \Xi_\text{eff} q \cos \left( \frac{\theta_{\bk, \bk + \bq}}{2} \right) 
\end{equation}
where $\Xi_\text{eff}$ is the effective deformation potential and the angular
part corresponds to the bare spinor overlap $\braket{\chi_{\bk +
    \bq}}{\chi_\bk}$ of the electronic wave function. In contrast to the more
complex angular dependence of the coupling matrix element predicted by the full
microscopic treatment presented here, the angular dependence of the
\emph{effective} coupling matrix element above suppresses only backscattering.
In the high-temperature regime where equipartitioning of the acoustic phonons
$N_{\bq} \sim k_\text{B}T / \hbar\omega_{\bq}$ applies, the relaxation time and
the resistivity take the following simples forms~\cite{Sarma:Acoustic}
\begin{equation}
  \label{eq:highT}
  \frac{1}{\tau_\bk} = \frac{1}{\hbar^3} 
                       \frac{\Xi_\text{eff}^2 k_\text{B} T}
                       {4\rho v_F^2 c^2} \varepsilon_\bk 
  \quad ; \quad
  \rho = \frac{\pi \Xi_\text{eff}^2 k_\text{B}T}{4e^2\hbar\rho v_F^2 c^2} ,
\end{equation}
where the factor of $4$ in the denominators stems from the chiral nature of the
carriers through the assumed form of the coupling matrix element in
Eq.~\eqref{eq:M_Sarma}. These expressions are used almost exclusively to extract
the value of the \emph{effective} acoustic deformation potential in experimental
situations~\cite{Fuhrer:GrapheneOnSiO2,Kim:Suspended,Zhu:Deposition,Hone:BNSubstrate,Kim:Controlling}.

\section{Results}

In the following, we study the intrinsic acoustic phonon-limited mobility of
graphene in both the BG and linear resistivity regime using the full coupling
matrix elements as given by Eqs.~\eqref{eq:M_TA_abinitio}
and~\eqref{eq:M_LA_abinitio}. This allows us to establish the value of the
intrinsic \emph{effective} acoustic deformation potential in graphene. 
\begin{figure}[!t]
  \includegraphics[width=0.79\linewidth]{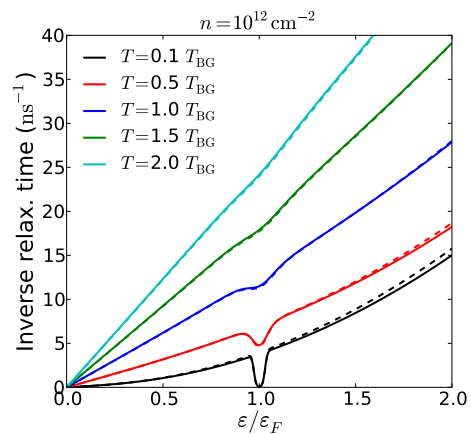}
  \caption{(Color online) Inverse relaxation time (valley-averaged) for acoustic
    phonon scattering on the TA and LA phonon in the BG regime at
    $n=10^{12}$~cm$^{-2}$ ($T_\text{BG} \approx 57$~K). The full lines show the
    results obtained with the full matrix elements given in
    Eqs.~\eqref{eq:M_TA_abinitio} and~\eqref{eq:M_LA_abinitio} and coupling
    constants extracted from \emph{ab-initio} calculations. The dashed lines
    show the result obtained with scattering on a single effective acoustic
    phonon with the coupling matrix element given by Eq.~\eqref{eq:M_Sarma}, a
    deformation potential of $\Xi_\text{eff} = 6.8$~eV and sound velocity
    $c=20\times10^3$~m/s.}
\label{fig:tau_vs_e}
\end{figure}

In Fig.~\ref{fig:tau_vs_e} we show the inverse of the valley-averaged relaxation
time as a function of energy for different temperatures and a carrier density of
$n=10^{12}$~cm$^{-2}$ corresponding to $\varepsilon_F \sim 117$~meV and
$T_\text{BG} \approx 57$~K for the LA phonon. Above the BG temperature, the
inverse relaxation time has the linear energy dependence of Eq.~\eqref{eq:highT}
and a slope proportional to the temperature. As the temperature is decreased
below $T_\text{BG}$, the freezing out of short wavelength phonons and the
sharpening of the Fermi surface result in limited phase space for phonon
scattering and an increased life time of the carriers at the Fermi energy. In
the expression for the relaxation time in Eq.~\eqref{eq:tau_acoustic} this
effect is accounted for by the Fermi and Bose distribution functions.  The
limited phase space available for phonon scattering, manifests itself in the
characteristic dip at the Fermi energy that evolves in the inverse relaxation
time with decreasing temperature~\cite{Sarma:Acoustic}. For all temperatures,
the linear energy dependence of the high-temperature result in
Eq.~\eqref{eq:highT} is recovered in the $\varepsilon \rightarrow 0$ limit.

By inspecting the individual contributions, we find that the inverse relaxation
time to a large extent is dominated by the contribution from the TA phonon both
in the high-temperature and the BG regime. The domination of the TA phonon can
be attributed to a number of factors. From Eq.~\eqref{eq:highT} it follows
directly that the lower sound velocity of the TA phonon leads to higher
scattering rate. Also related to the sound velocity is the lower BG temperature
of TA phonon which allows for full backscattering below the BG temperature of
the LA phonon. Secondly, the coupling matrix element for the TA phonon allows
for backscattering which is suppressed for the LA phonon. In the BG regime, the
domination of the TA phonon stems from the suppression of the coupling matrix
element for the LA phonon in the long-wavelength limit (see
Fig.~\ref{fig:couplings}). The observed dominance of the TA phonon is in
contrast to the often used assumption that only the LA phonon couples to charge
carriers in graphene~\cite{Sarma:Acoustic}.

In order to determine the intrinsic value of the \emph{effective} deformation
potential in graphene, we also calculate the relaxation time using coupling
matrix element in Eq.~\eqref{eq:M_Sarma}. The dashed lines in
Fig.~\ref{fig:tau_vs_e} show the inverse relaxation time calculated with an
\emph{effective} deformation potential and sound velocity of $6.8$~eV and $20.0
\times 10^3$~m/s, respectively. It is seen to reproduce the relaxation time
based on the full matrix elements very well for the energy range shown. While
the extracted value for the acoustic deformation potential is much smaller than
experimental
values~\cite{Fuhrer:GrapheneOnSiO2,Kim:Suspended,Zhu:Deposition,Hone:BNSubstrate,Kim:Controlling},
it is in better agreement with recently reported \emph{ab-initio} results
yielding $4.5$~eV~\cite{Kim:ElPhGraphene}.

\begin{figure}[!t]
  \begin{minipage}{1.0\linewidth}
    \includegraphics[width=0.69\linewidth]{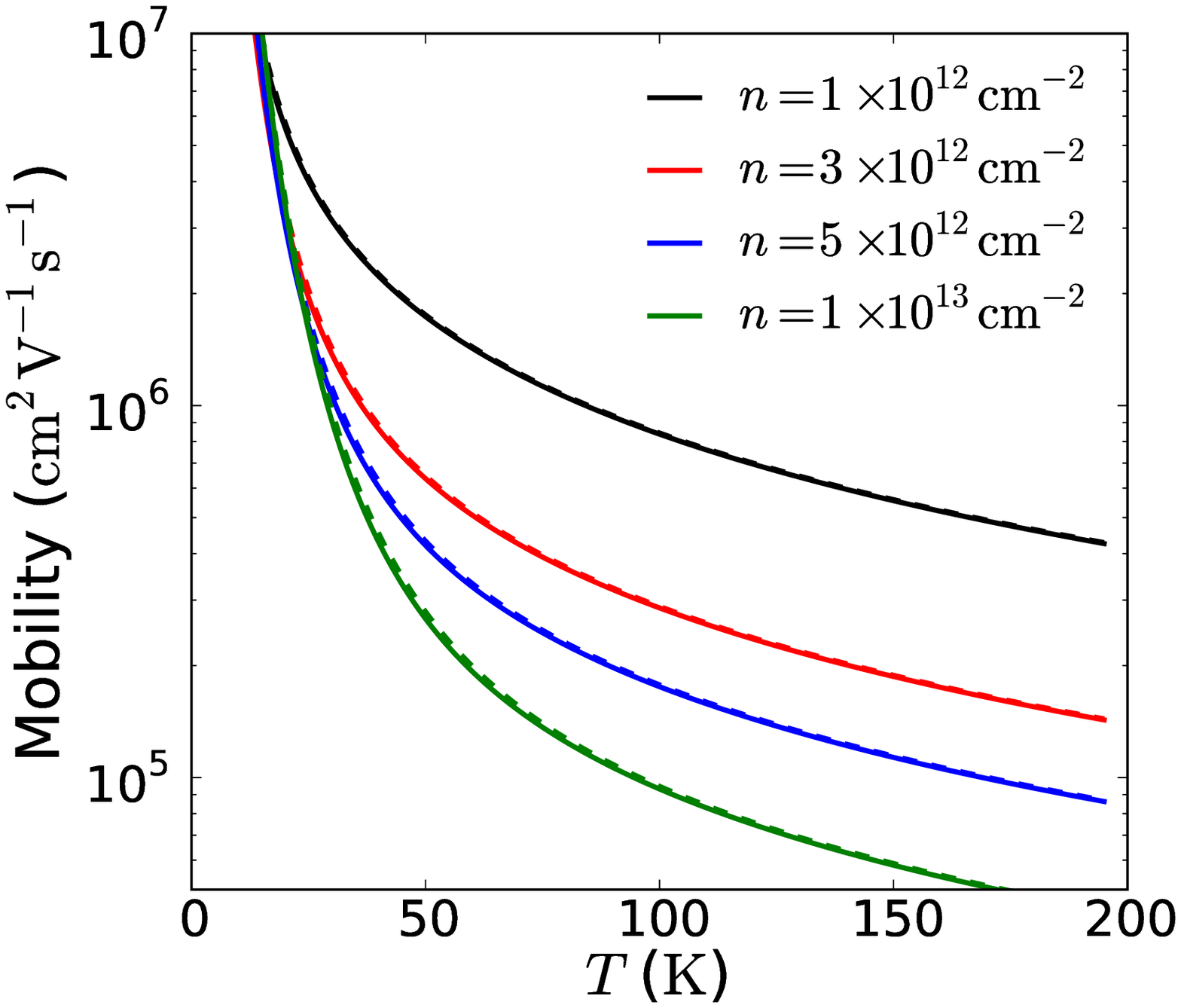}
    \includegraphics[width=0.69\linewidth]{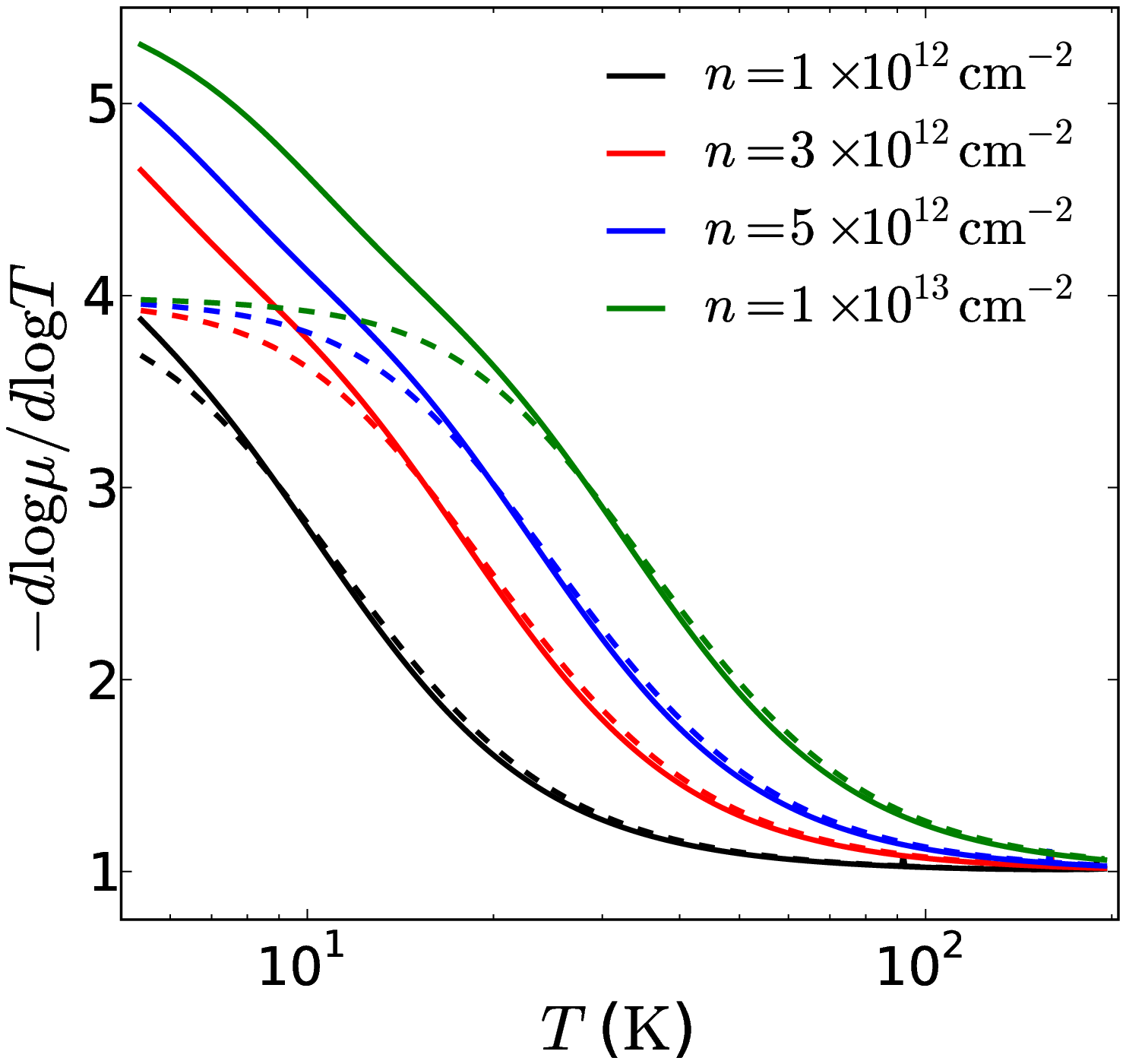}
  \end{minipage}
  \caption{(Color online) Mobility vs temperature for carrier densities $10^{12}
    - 10^{13}$~cm$^{-2}$ (upper plot). The lower plot shows the temperature
    dependence of the exponent $\alpha$ in the temperature dependence $\mu \sim
    T^{-\alpha}$ of the mobility.}
\label{fig:mobility}
\end{figure}
Figure~\ref{fig:mobility} summarizes the calculated acoustic phonon-limited
mobility as a function of temperature for carrier densities $10^{12} -
10^{13}$~cm$^{-2}$. The mobility calculated with the above-mentioned
\emph{effective} coupling parameters (dashed lines) reproduces the full
calculation extremely well. For all carrier densities the mobility shows a
transition from the linear $\mu \sim T^{-1}$ high-temperature behavior to a more
pronounced $\mu \sim T^{-\alpha}$ temperature dependence with $\alpha > 1$ in
the BG regime. At temperatures $T> T_\text{BG}$, the decrease in the mobility
with increasing carrier density stems from the linear energy dependence of the
density of states which provides more phase space for phonon scattering for
higher Fermi levels. At low temperatures $T < T_\text{BG}$ where scattering on
the full Fermi surface is frozen out, the local value of the coupling matrix
element becomes important and the mobilities for the different carrier densities
approach a common value.

The right plot shows the temperature dependence of the exponent $\alpha$ for the
same set of carrier densities which correspond to BG temperatures $T_\text{BG}
\sim 57-180$~K for the LA phonon. From this plot it is more clear that the
departure away from the linear temperature dependence happens for $T \sim
T_\text{BG}$. For $T < T_\text{BG}$, the exponent increases monotonically.
Surprisingly, the exponents obtained from the mobility calculated with the full
coupling matrix elements do not saturate at $\alpha = 4$ as predicted by the
\emph{effective} coupling matrix element (dashed
lines)~\cite{Sarma:Acoustic,Sarma:Chirality}. Thus, even in the absence of
screening, the mobility of graphene should take on a temperature dependence with
$\alpha > 4$ at sufficiently low temperatures. With carrier screening taken into
account, this behavior is reinforced~\cite{Sarma:Chirality}

The purely intrinsic mobilities calculated here are significantly higher than
previously reported theoretical values~\cite{Sarma:Acoustic}. This is reflected
directly in the extracted deformation potential parameter which is considerably
lower than commonly used values. For a carrier density of $n= 10^{12}$~cm$^{-2}$
a room-temperature mobility in excess of $10^5$~cm$^2$~V$^{-1}$~s$^{-1}$ is here
predicted. The associated scattering rate which is given by the relaxation time
in Eq.~\eqref{eq:highT} in the high-temperature regime, is from
Fig.~\ref{fig:tau_vs_e} estimated to be $\tau^{-1} \sim 10^{11}$~s$^{-1}$ at the
Fermi level corresponding to a mean-free path of $\lambda = v_F / \tau \sim
1000$~nm. Such an extremely large mean-free path may open the opportunity to
study coherent transport in relatively large graphene structures. We note that
the quasiparticle scattering rate observable in photoemission spectroscopy
(ARPES) develops a similar dip at the Fermi level for $T < T_\text{BG}$
resulting in long-lived quasiparticles.

\section{Conclusions and discussions}

In the present study the acoustic electron-phonon interaction in graphene has
been analyzed in detail. The exact analytic forms of the coupling matrix
elements in long-wavelength limit were found match the calculated
first-principles matrix elements almost quantitatively even at shorter
wavelengths. As previously predicted~\cite{Sarma:Acoustic,Sarma:Chirality}, the
calculated mobilities showed a transition from a $\mu \sim T^{-1}$ to a $\mu
\sim T^{-\alpha}$ temperature dependence with $\alpha > 1$ below the BG
temperature. However, contrary to earlier studies we found that the full
coupling matrix elements cause the temperature dependence of the mobility to
increase beyond $\alpha = 4$ which is otherwise only observed when screening is
included~\cite{Sarma:Chirality}.

By fitting the results with an \emph{effective} acoustic phonon the acoustic
deformation potential is found to be $\Xi_\text{eff} = 6.8$~eV. Since this is
much lower than the experimentally determined values $18-29$~eV, our results
suggest that the acoustic phonon-limited transport in substrate-supported
graphene is at present not fully understood. Possible explanations for the large
experimental deformation potentials could be (i) substrate-induced modifications
of the band structure~\cite{Louie:New,Brink:Incommensurate} that modifies the
chirality (and thereby the angular dependence of the coupling matrix element) of
the electronic states and/or the Fermi velocity, and (ii) the existence of
additional acoustic phonons not considered in the present work. This could be
either the intrinsic flexural phonon which may be modified when graphene is
lying on a substrate or surface-acoustic phonons on the substrate which have
previously been studied in 2DEGs~\cite{Knabchen:SurfaceAcoustic}. The relatively
large variations in the experimental deformation potentials also indicate that
the deformation potential depends on experimental factors such as e.g. the
substrate.

\begin{acknowledgments}
  The authors would like to thank T.~Markussen and A.-P.~Jauho for useful
  comments on the manuscript. KK has been partially supported by the Center on
  Nanostructuring for Efficient Energy Conversion (CNEEC) at Stanford
  University, an Energy Frontier Research Center funded by the U.S. Department
  of Energy, Office of Science, Office of Basic Energy Sciences under Award
  Number DE-SC0001060. CAMD is supported by the Lundbeck Foundation.
\end{acknowledgments}


\begin{thebibliography}{42}
\expandafter\ifx\csname natexlab\endcsname\relax\def\natexlab#1{#1}\fi
\expandafter\ifx\csname bibnamefont\endcsname\relax
  \def\bibnamefont#1{#1}\fi
\expandafter\ifx\csname bibfnamefont\endcsname\relax
  \def\bibfnamefont#1{#1}\fi
\expandafter\ifx\csname citenamefont\endcsname\relax
  \def\citenamefont#1{#1}\fi
\expandafter\ifx\csname url\endcsname\relax
  \def\url#1{\texttt{#1}}\fi
\expandafter\ifx\csname urlprefix\endcsname\relax\def\urlprefix{URL }\fi
\providecommand{\bibinfo}[2]{#2}
\providecommand{\eprint}[2][]{\url{#2}}

\bibitem[{\citenamefont{Novoselov et~al.}(2004)\citenamefont{Novoselov, Geim,
  Morozov, Jiang, Zhang, Dubonos, Grigorieva, and Firsov}}]{Firsov:Science}
\bibinfo{author}{\bibfnamefont{K.~S.} \bibnamefont{Novoselov}},
  \bibinfo{author}{\bibfnamefont{A.~K.} \bibnamefont{Geim}},
  \bibinfo{author}{\bibfnamefont{S.~V.} \bibnamefont{Morozov}},
  \bibinfo{author}{\bibfnamefont{D.}~\bibnamefont{Jiang}},
  \bibinfo{author}{\bibfnamefont{Y.}~\bibnamefont{Zhang}},
  \bibinfo{author}{\bibfnamefont{S.~V.} \bibnamefont{Dubonos}},
  \bibinfo{author}{\bibfnamefont{I.~V.} \bibnamefont{Grigorieva}},
  \bibnamefont{and} \bibinfo{author}{\bibfnamefont{A.~A.}
  \bibnamefont{Firsov}}, \bibinfo{journal}{Science}
  \textbf{\bibinfo{volume}{306}}, \bibinfo{pages}{666} (\bibinfo{year}{2004}).

\bibitem[{\citenamefont{Geim and Novoselov}(2007)}]{Geim:Graphene}
\bibinfo{author}{\bibfnamefont{A.~K.} \bibnamefont{Geim}} \bibnamefont{and}
  \bibinfo{author}{\bibfnamefont{K.~S.} \bibnamefont{Novoselov}},
  \bibinfo{journal}{Nature Mat.} \textbf{\bibinfo{volume}{6}},
  \bibinfo{pages}{183} (\bibinfo{year}{2007}).

\bibitem[{\citenamefont{Neto et~al.}(2009)\citenamefont{Neto, Guinea, Peres,
  Novoselov, and Geim}}]{RMP:Graphene}
\bibinfo{author}{\bibfnamefont{A.~H.~C.} \bibnamefont{Neto}},
  \bibinfo{author}{\bibfnamefont{F.}~\bibnamefont{Guinea}},
  \bibinfo{author}{\bibfnamefont{N.~M.~R.} \bibnamefont{Peres}},
  \bibinfo{author}{\bibfnamefont{K.~S.} \bibnamefont{Novoselov}},
  \bibnamefont{and} \bibinfo{author}{\bibfnamefont{A.~K.} \bibnamefont{Geim}},
  \bibinfo{journal}{Rev. Mod. Phys.} \textbf{\bibinfo{volume}{81}},
  \bibinfo{pages}{109} (\bibinfo{year}{2009}).

\bibitem[{\citenamefont{{Das Sarma} et~al.}(2011)\citenamefont{{Das Sarma},
  Adam, Hwang, and Rossi}}]{Sarma:RMP}
\bibinfo{author}{\bibfnamefont{S.}~\bibnamefont{{Das Sarma}}},
  \bibinfo{author}{\bibfnamefont{S.}~\bibnamefont{Adam}},
  \bibinfo{author}{\bibfnamefont{E.~H.} \bibnamefont{Hwang}}, \bibnamefont{and}
  \bibinfo{author}{\bibfnamefont{E.}~\bibnamefont{Rossi}},
  \bibinfo{journal}{Rev. Mod. Phys.} \textbf{\bibinfo{volume}{83}},
  \bibinfo{pages}{407} (\bibinfo{year}{2011}).

\bibitem[{\citenamefont{Chen et~al.}(2008)\citenamefont{Chen, Jang, Xiao,
  Ishigami, and Fuhrer}}]{Fuhrer:GrapheneOnSiO2}
\bibinfo{author}{\bibfnamefont{J.-H.} \bibnamefont{Chen}},
  \bibinfo{author}{\bibfnamefont{C.}~\bibnamefont{Jang}},
  \bibinfo{author}{\bibfnamefont{S.}~\bibnamefont{Xiao}},
  \bibinfo{author}{\bibfnamefont{M.}~\bibnamefont{Ishigami}}, \bibnamefont{and}
  \bibinfo{author}{\bibfnamefont{M.~S.} \bibnamefont{Fuhrer}},
  \bibinfo{journal}{Nature Nano.} \textbf{\bibinfo{volume}{3}},
  \bibinfo{pages}{206} (\bibinfo{year}{2008}).

\bibitem[{\citenamefont{Hwang and {Das Sarma}}(2009)}]{Sarma:ScreeningInduced}
\bibinfo{author}{\bibfnamefont{E.~H.} \bibnamefont{Hwang}} \bibnamefont{and}
  \bibinfo{author}{\bibfnamefont{S.}~\bibnamefont{{Das Sarma}}},
  \bibinfo{journal}{Phys. Rev. B} \textbf{\bibinfo{volume}{79}},
  \bibinfo{pages}{165404} (\bibinfo{year}{2009}).

\bibitem[{\citenamefont{Efetov and Kim}(2010)}]{Kim:Controlling}
\bibinfo{author}{\bibfnamefont{D.~K.} \bibnamefont{Efetov}} \bibnamefont{and}
  \bibinfo{author}{\bibfnamefont{P.}~\bibnamefont{Kim}},
  \bibinfo{journal}{Phys. Rev. Lett.} \textbf{\bibinfo{volume}{105}},
  \bibinfo{pages}{256805} (\bibinfo{year}{2010}).

\bibitem[{\citenamefont{Dean et~al.}(2010)\citenamefont{Dean, Young, Meric,
  Lee, Wang, Sorgenfrei, Watanabe, Taniguchi, Kim, Shepard
  et~al.}}]{Hone:BNSubstrate}
\bibinfo{author}{\bibfnamefont{C.~R.} \bibnamefont{Dean}},
  \bibinfo{author}{\bibfnamefont{A.~F.} \bibnamefont{Young}},
  \bibinfo{author}{\bibfnamefont{I.}~\bibnamefont{Meric}},
  \bibinfo{author}{\bibfnamefont{C.}~\bibnamefont{Lee}},
  \bibinfo{author}{\bibfnamefont{L.}~\bibnamefont{Wang}},
  \bibinfo{author}{\bibfnamefont{S.}~\bibnamefont{Sorgenfrei}},
  \bibinfo{author}{\bibfnamefont{K.}~\bibnamefont{Watanabe}},
  \bibinfo{author}{\bibfnamefont{T.}~\bibnamefont{Taniguchi}},
  \bibinfo{author}{\bibfnamefont{P.}~\bibnamefont{Kim}},
  \bibinfo{author}{\bibfnamefont{K.~L.} \bibnamefont{Shepard}},
  \bibnamefont{et~al.}, \bibinfo{journal}{Nature Nano.}
  \textbf{\bibinfo{volume}{5}}, \bibinfo{pages}{722} (\bibinfo{year}{2010}).

\bibitem[{\citenamefont{Zou et~al.}(2010)\citenamefont{Zou, Hong, Keefer, and
  Zhu}}]{Zhu:Deposition}
\bibinfo{author}{\bibfnamefont{K.}~\bibnamefont{Zou}},
  \bibinfo{author}{\bibfnamefont{X.}~\bibnamefont{Hong}},
  \bibinfo{author}{\bibfnamefont{D.}~\bibnamefont{Keefer}}, \bibnamefont{and}
  \bibinfo{author}{\bibfnamefont{J.}~\bibnamefont{Zhu}},
  \bibinfo{journal}{Phys. Rev. Lett.} \textbf{\bibinfo{volume}{105}},
  \bibinfo{pages}{126601} (\bibinfo{year}{2010}).

\bibitem[{\citenamefont{Pachoud et~al.}(2010)\citenamefont{Pachoud, Jaiswal,
  Ang, Loh, and {\"O}zyilmaz}}]{Ozyilmaz:ElectrolyteGate}
\bibinfo{author}{\bibfnamefont{A.}~\bibnamefont{Pachoud}},
  \bibinfo{author}{\bibfnamefont{M.}~\bibnamefont{Jaiswal}},
  \bibinfo{author}{\bibfnamefont{P.~K.} \bibnamefont{Ang}},
  \bibinfo{author}{\bibfnamefont{K.~P.} \bibnamefont{Loh}}, \bibnamefont{and}
  \bibinfo{author}{\bibfnamefont{B.}~\bibnamefont{{\"O}zyilmaz}},
  \bibinfo{journal}{Europhys. Lett.} \textbf{\bibinfo{volume}{92}},
  \bibinfo{pages}{27001} (\bibinfo{year}{2010}).

\bibitem[{\citenamefont{Bolotin et~al.}(2008)\citenamefont{Bolotin, Sikes,
  Hone, Stormer, and Kim}}]{Kim:Suspended}
\bibinfo{author}{\bibfnamefont{K.~I.} \bibnamefont{Bolotin}},
  \bibinfo{author}{\bibfnamefont{K.~J.} \bibnamefont{Sikes}},
  \bibinfo{author}{\bibfnamefont{J.}~\bibnamefont{Hone}},
  \bibinfo{author}{\bibfnamefont{H.~L.} \bibnamefont{Stormer}},
  \bibnamefont{and} \bibinfo{author}{\bibfnamefont{P.}~\bibnamefont{Kim}},
  \bibinfo{journal}{Phys. Rev. Lett.} \textbf{\bibinfo{volume}{101}},
  \bibinfo{pages}{096802} (\bibinfo{year}{2008}).

\bibitem[{\citenamefont{Xue et~al.}(2011)\citenamefont{Xue, Sanchez-Yamagishi,
  Bulmash, Jacquod, Deshpande, Watanabe, Taniguchi, Jarillo-Herrero, and
  LeRoy}}]{LeRoy:STMhBN}
\bibinfo{author}{\bibfnamefont{J.}~\bibnamefont{Xue}},
  \bibinfo{author}{\bibfnamefont{J.}~\bibnamefont{Sanchez-Yamagishi}},
  \bibinfo{author}{\bibfnamefont{D.}~\bibnamefont{Bulmash}},
  \bibinfo{author}{\bibfnamefont{P.}~\bibnamefont{Jacquod}},
  \bibinfo{author}{\bibfnamefont{A.}~\bibnamefont{Deshpande}},
  \bibinfo{author}{\bibfnamefont{K.}~\bibnamefont{Watanabe}},
  \bibinfo{author}{\bibfnamefont{T.}~\bibnamefont{Taniguchi}},
  \bibinfo{author}{\bibfnamefont{P.}~\bibnamefont{Jarillo-Herrero}},
  \bibnamefont{and} \bibinfo{author}{\bibfnamefont{B.~J.} \bibnamefont{LeRoy}},
  \bibinfo{journal}{Nature Mat.} \textbf{\bibinfo{volume}{10}},
  \bibinfo{pages}{282} (\bibinfo{year}{2011}).

\bibitem[{\citenamefont{Decker et~al.}(2011)\citenamefont{Decker, Wang, Brar,
  Regan, Tsai, Wu, Gannett, Zettl, and Crommie}}]{Crommie:LocalBN}
\bibinfo{author}{\bibfnamefont{R.}~\bibnamefont{Decker}},
  \bibinfo{author}{\bibfnamefont{Y.}~\bibnamefont{Wang}},
  \bibinfo{author}{\bibfnamefont{V.~W.} \bibnamefont{Brar}},
  \bibinfo{author}{\bibfnamefont{W.}~\bibnamefont{Regan}},
  \bibinfo{author}{\bibfnamefont{H.-Z.} \bibnamefont{Tsai}},
  \bibinfo{author}{\bibfnamefont{Q.}~\bibnamefont{Wu}},
  \bibinfo{author}{\bibfnamefont{W.}~\bibnamefont{Gannett}},
  \bibinfo{author}{\bibfnamefont{A.}~\bibnamefont{Zettl}}, \bibnamefont{and}
  \bibinfo{author}{\bibfnamefont{M.~F.} \bibnamefont{Crommie}},
  \bibinfo{journal}{Nano. Lett.} \textbf{\bibinfo{volume}{11}},
  \bibinfo{pages}{2291} (\bibinfo{year}{2011}).

\bibitem[{\citenamefont{Dean et~al.}(2011)\citenamefont{Dean, Young,
  {Cadden-Zimansky}, Wang, Ren, Watanabe, Taniguchi, Kim, Hone, and
  Shepard}}]{Shepard:HallEffect}
\bibinfo{author}{\bibfnamefont{C.~R.} \bibnamefont{Dean}},
  \bibinfo{author}{\bibfnamefont{A.~F.} \bibnamefont{Young}},
  \bibinfo{author}{\bibfnamefont{P.}~\bibnamefont{{Cadden-Zimansky}}},
  \bibinfo{author}{\bibfnamefont{L.}~\bibnamefont{Wang}},
  \bibinfo{author}{\bibfnamefont{H.}~\bibnamefont{Ren}},
  \bibinfo{author}{\bibfnamefont{K.}~\bibnamefont{Watanabe}},
  \bibinfo{author}{\bibfnamefont{T.}~\bibnamefont{Taniguchi}},
  \bibinfo{author}{\bibfnamefont{P.}~\bibnamefont{Kim}},
  \bibinfo{author}{\bibfnamefont{J.}~\bibnamefont{Hone}}, \bibnamefont{and}
  \bibinfo{author}{\bibfnamefont{K.~L.} \bibnamefont{Shepard}},
  \bibinfo{journal}{Nature Phys.} \textbf{\bibinfo{volume}{7}},
  \bibinfo{pages}{693} (\bibinfo{year}{2011}).

\bibitem[{\citenamefont{Zomer et~al.}(2011)\citenamefont{Zomer, Dash, Tombros,
  and van Wees}}]{Wees:Transfer}
\bibinfo{author}{\bibfnamefont{P.~J.} \bibnamefont{Zomer}},
  \bibinfo{author}{\bibfnamefont{S.~P.} \bibnamefont{Dash}},
  \bibinfo{author}{\bibfnamefont{N.}~\bibnamefont{Tombros}}, \bibnamefont{and}
  \bibinfo{author}{\bibfnamefont{B.~J.} \bibnamefont{van Wees}},
  \bibinfo{journal}{Appl. Phys. Lett.} \textbf{\bibinfo{volume}{99}},
  \bibinfo{pages}{232104} (\bibinfo{year}{2011}).

\bibitem[{\citenamefont{Fratini and Guinea}(2008)}]{Guinea:SubstrateLimited}
\bibinfo{author}{\bibfnamefont{S.}~\bibnamefont{Fratini}} \bibnamefont{and}
  \bibinfo{author}{\bibfnamefont{F.}~\bibnamefont{Guinea}},
  \bibinfo{journal}{Phys. Rev. B} \textbf{\bibinfo{volume}{77}},
  \bibinfo{pages}{195415} (\bibinfo{year}{2008}).

\bibitem[{\citenamefont{Konar et~al.}(2010)\citenamefont{Konar, Fang, and
  Jena}}]{Jena:High-k}
\bibinfo{author}{\bibfnamefont{A.}~\bibnamefont{Konar}},
  \bibinfo{author}{\bibfnamefont{T.}~\bibnamefont{Fang}}, \bibnamefont{and}
  \bibinfo{author}{\bibfnamefont{D.}~\bibnamefont{Jena}},
  \bibinfo{journal}{Phys. Rev. B} \textbf{\bibinfo{volume}{82}},
  \bibinfo{pages}{115452} (\bibinfo{year}{2010}).

\bibitem[{\citenamefont{Li et~al.}(2010)\citenamefont{Li, Barry, Zavada,
  Nardelli, and Kim}}]{Kim:SurfacePolar}
\bibinfo{author}{\bibfnamefont{X.}~\bibnamefont{Li}},
  \bibinfo{author}{\bibfnamefont{E.~A.} \bibnamefont{Barry}},
  \bibinfo{author}{\bibfnamefont{J.~M.} \bibnamefont{Zavada}},
  \bibinfo{author}{\bibfnamefont{M.~B.} \bibnamefont{Nardelli}},
  \bibnamefont{and} \bibinfo{author}{\bibfnamefont{K.~W.} \bibnamefont{Kim}},
  \bibinfo{journal}{Appl. Phys. Lett.} \textbf{\bibinfo{volume}{97}},
  \bibinfo{pages}{232105} (\bibinfo{year}{2010}).

\bibitem[{\citenamefont{Ziman}(1960)}]{Ziman}
\bibinfo{author}{\bibfnamefont{J.~M.} \bibnamefont{Ziman}},
  \emph{\bibinfo{title}{Electrons and phonons}} (\bibinfo{publisher}{Oxford
  University Press}, \bibinfo{address}{London}, \bibinfo{year}{1960}).

\bibitem[{\citenamefont{Hwang and {Das Sarma}}(2008)}]{Sarma:Acoustic}
\bibinfo{author}{\bibfnamefont{E.~H.} \bibnamefont{Hwang}} \bibnamefont{and}
  \bibinfo{author}{\bibfnamefont{S.}~\bibnamefont{{Das Sarma}}},
  \bibinfo{journal}{Phys. Rev. B} \textbf{\bibinfo{volume}{77}},
  \bibinfo{pages}{115449} (\bibinfo{year}{2008}).

\bibitem[{\citenamefont{Min et~al.}(2011)\citenamefont{Min, Hwang, and {Das
  Sarma}}}]{Sarma:Chirality}
\bibinfo{author}{\bibfnamefont{H.}~\bibnamefont{Min}},
  \bibinfo{author}{\bibfnamefont{E.~H.} \bibnamefont{Hwang}}, \bibnamefont{and}
  \bibinfo{author}{\bibfnamefont{S.}~\bibnamefont{{Das Sarma}}},
  \bibinfo{journal}{Phys. Rev. B} \textbf{\bibinfo{volume}{83}},
  \bibinfo{pages}{161404} (\bibinfo{year}{2011}).

\bibitem[{\citenamefont{Perebeinos and Avouris}(2010)}]{Avouris:Inelastic}
\bibinfo{author}{\bibfnamefont{V.}~\bibnamefont{Perebeinos}} \bibnamefont{and}
  \bibinfo{author}{\bibfnamefont{P.}~\bibnamefont{Avouris}},
  \bibinfo{journal}{Phys. Rev. B} \textbf{\bibinfo{volume}{81}},
  \bibinfo{pages}{195442} (\bibinfo{year}{2010}).

\bibitem[{\citenamefont{Borysenko et~al.}(2010)\citenamefont{Borysenko, Mullen,
  Barry, Paul, Semenov, Zavada, Nardelli, and Kim}}]{Kim:ElPhGraphene}
\bibinfo{author}{\bibfnamefont{K.~M.} \bibnamefont{Borysenko}},
  \bibinfo{author}{\bibfnamefont{J.~T.} \bibnamefont{Mullen}},
  \bibinfo{author}{\bibfnamefont{E.~A.} \bibnamefont{Barry}},
  \bibinfo{author}{\bibfnamefont{S.}~\bibnamefont{Paul}},
  \bibinfo{author}{\bibfnamefont{Y.~G.} \bibnamefont{Semenov}},
  \bibinfo{author}{\bibfnamefont{J.~M.} \bibnamefont{Zavada}},
  \bibinfo{author}{\bibfnamefont{M.~B.} \bibnamefont{Nardelli}},
  \bibnamefont{and} \bibinfo{author}{\bibfnamefont{K.~W.} \bibnamefont{Kim}},
  \bibinfo{journal}{Phys. Rev. B} \textbf{\bibinfo{volume}{81}},
  \bibinfo{pages}{121412} (\bibinfo{year}{2010}).

\bibitem[{\citenamefont{Shishir and Ferry}(2009)}]{Ferry:Intrinsic}
\bibinfo{author}{\bibfnamefont{R.~S.} \bibnamefont{Shishir}} \bibnamefont{and}
  \bibinfo{author}{\bibfnamefont{D.~K.} \bibnamefont{Ferry}},
  \bibinfo{journal}{J. Phys.: Condens. Matter} \textbf{\bibinfo{volume}{21}},
  \bibinfo{pages}{232204} (\bibinfo{year}{2009}).

\bibitem[{\citenamefont{Stauber et~al.}(2007)\citenamefont{Stauber, Peres, and
  Guinea}}]{Guinea:ElectronicTransport}
\bibinfo{author}{\bibfnamefont{T.}~\bibnamefont{Stauber}},
  \bibinfo{author}{\bibfnamefont{N.~M.~R.} \bibnamefont{Peres}},
  \bibnamefont{and} \bibinfo{author}{\bibfnamefont{F.}~\bibnamefont{Guinea}},
  \bibinfo{journal}{Phys. Rev. B} \textbf{\bibinfo{volume}{76}},
  \bibinfo{pages}{205423} (\bibinfo{year}{2007}).

\bibitem[{\citenamefont{Woods and Mahan}(2000)}]{Mahan:ElphGraphene}
\bibinfo{author}{\bibfnamefont{L.~M.} \bibnamefont{Woods}} \bibnamefont{and}
  \bibinfo{author}{\bibfnamefont{G.~D.} \bibnamefont{Mahan}},
  \bibinfo{journal}{Phys. Rev. B} \textbf{\bibinfo{volume}{61}},
  \bibinfo{pages}{10651} (\bibinfo{year}{2000}).

\bibitem[{\citenamefont{Castro et~al.}(2010)\citenamefont{Castro, Ochoa,
  Katsnelson, Gorbachev, Elias, Novoselov, Geim, and Guinea}}]{Guinea:Flexural}
\bibinfo{author}{\bibfnamefont{E.~V.} \bibnamefont{Castro}},
  \bibinfo{author}{\bibfnamefont{H.}~\bibnamefont{Ochoa}},
  \bibinfo{author}{\bibfnamefont{M.~I.} \bibnamefont{Katsnelson}},
  \bibinfo{author}{\bibfnamefont{R.~V.} \bibnamefont{Gorbachev}},
  \bibinfo{author}{\bibfnamefont{D.~C.} \bibnamefont{Elias}},
  \bibinfo{author}{\bibfnamefont{K.~S.} \bibnamefont{Novoselov}},
  \bibinfo{author}{\bibfnamefont{A.~K.} \bibnamefont{Geim}}, \bibnamefont{and}
  \bibinfo{author}{\bibfnamefont{F.}~\bibnamefont{Guinea}},
  \bibinfo{journal}{Phys. Rev. Lett.} \textbf{\bibinfo{volume}{105}},
  \bibinfo{pages}{266601} (\bibinfo{year}{2010}).

\bibitem[{\citenamefont{Kaasbjerg et~al.}(2012)\citenamefont{Kaasbjerg,
  Thygesen, and Jacobsen}}]{Kaasbjerg:MoS2}
\bibinfo{author}{\bibfnamefont{K.}~\bibnamefont{Kaasbjerg}},
  \bibinfo{author}{\bibfnamefont{K.~S.} \bibnamefont{Thygesen}},
  \bibnamefont{and} \bibinfo{author}{\bibfnamefont{K.~W.}
  \bibnamefont{Jacobsen}}, \bibinfo{journal}{Phys. Rev. B}
  (\bibinfo{year}{2012}), \bibinfo{note}{submitted}.

\bibitem[{\citenamefont{Enkovaara et~al.}(2010)\citenamefont{Enkovaara,
  Rostgaard, Mortensen, Chen, Dulak, Ferrighi, Gavnholt, Glinsvad, Haikola,
  Hansen et~al.}}]{GPAW}
\bibinfo{author}{\bibfnamefont{J.~.} \bibnamefont{Enkovaara}},
  \bibinfo{author}{\bibfnamefont{C.}~\bibnamefont{Rostgaard}},
  \bibinfo{author}{\bibfnamefont{J.~J.} \bibnamefont{Mortensen}},
  \bibinfo{author}{\bibfnamefont{J.}~\bibnamefont{Chen}},
  \bibinfo{author}{\bibfnamefont{M.}~\bibnamefont{Dulak}},
  \bibinfo{author}{\bibfnamefont{L.}~\bibnamefont{Ferrighi}},
  \bibinfo{author}{\bibfnamefont{J.}~\bibnamefont{Gavnholt}},
  \bibinfo{author}{\bibfnamefont{C.}~\bibnamefont{Glinsvad}},
  \bibinfo{author}{\bibfnamefont{V.}~\bibnamefont{Haikola}},
  \bibinfo{author}{\bibfnamefont{H.~A.} \bibnamefont{Hansen}},
  \bibnamefont{et~al.}, \bibinfo{journal}{J. Phys.: Condens. Matter}
  \textbf{\bibinfo{volume}{22}}, \bibinfo{pages}{253202}
  (\bibinfo{year}{2010}).

\bibitem[{\citenamefont{Ma{\~n}es}(2007)}]{Manes:Symmetry}
\bibinfo{author}{\bibfnamefont{J.~L.} \bibnamefont{Ma{\~n}es}},
  \bibinfo{journal}{Phys. Rev. B} \textbf{\bibinfo{volume}{76}},
  \bibinfo{pages}{045430} (\bibinfo{year}{2007}).

\bibitem[{\citenamefont{Hwang et~al.}(2007)\citenamefont{Hwang, Adam, and {Das
  Sarma}}}]{Sarma:CarrierTransport}
\bibinfo{author}{\bibfnamefont{E.~H.} \bibnamefont{Hwang}},
  \bibinfo{author}{\bibfnamefont{S.}~\bibnamefont{Adam}}, \bibnamefont{and}
  \bibinfo{author}{\bibfnamefont{S.}~\bibnamefont{{Das Sarma}}},
  \bibinfo{journal}{Phys. Rev. Lett.} \textbf{\bibinfo{volume}{98}},
  \bibinfo{pages}{186806} (\bibinfo{year}{2007}).

\bibitem[{\citenamefont{Wunsch et~al.}(2006)\citenamefont{Wunsch, Stauber,
  Sols, and Guinea}}]{Guinea:Dynamical}
\bibinfo{author}{\bibfnamefont{B.}~\bibnamefont{Wunsch}},
  \bibinfo{author}{\bibfnamefont{T.}~\bibnamefont{Stauber}},
  \bibinfo{author}{\bibfnamefont{F.}~\bibnamefont{Sols}}, \bibnamefont{and}
  \bibinfo{author}{\bibfnamefont{F.}~\bibnamefont{Guinea}},
  \bibinfo{journal}{New J. Phys.} \textbf{\bibinfo{volume}{8}},
  \bibinfo{pages}{318} (\bibinfo{year}{2006}).

\bibitem[{\citenamefont{Hwang and {Das Sarma}}(2007)}]{Sarma:Dielectric}
\bibinfo{author}{\bibfnamefont{E.~H.} \bibnamefont{Hwang}} \bibnamefont{and}
  \bibinfo{author}{\bibfnamefont{S.}~\bibnamefont{{Das Sarma}}},
  \bibinfo{journal}{Phys. Rev. B} \textbf{\bibinfo{volume}{75}},
  \bibinfo{pages}{205418} (\bibinfo{year}{2007}).

\bibitem[{foo({\natexlab{a}})}]{footnote1}
\bibinfo{note}{Since $c_\lambda / v_F \sim 10^{-2}$ for graphene, the acoustic
  phonon energy can to a good approximation be neglected in argument of energy
  conserving $\delta$-functions of Eq.~\eqref{eq:P_acoustic}.}

\bibitem[{foo({\natexlab{b}})}]{footnote2}
\bibinfo{note}{For the numerical evaluation of Eq.~\eqref{eq:tau_acoustic}, the
  identity $$ f_{\bk} \left(1 - f_{\bk \pm \bq} \right) N_{\pm\bq} = f_{\bk \pm
  \bq} \left(1 - f_{\bk} \right) \left(1 + N_{\pm\bq} \right) , $$ where
  $f_{\bk\pm\bq} = f(\varepsilon_\bk \pm \hbar\omega_\bq)$ and $N_{\pm\bq} =
  N(\pm \hbar\omega_\bq)$ is understood, comes in very handy.}

\bibitem[{foo({\natexlab{c}})}]{footnote3}
\bibinfo{note}{The coupling matrix elements for the electron-phonon coupling
  have been calculated within DFT-LDA using a $7\times 7$ supercell and a DZP
  basis for the electronic Bloch states.}

\bibitem[{foo({\natexlab{d}})}]{footnote4}
\bibinfo{note}{At shorter wavelengths, the strictly transverse and longitudinal
  character of the modes vanishes. For the TA phonon this implies that the
  transverse character is only retained in certain high symmetry directions of
  the Brillouin zone. For the LA phonon we find that the mode vector retains
  its long-wavelength character for $q$-values up to $\sim 30\%$ of the
  Brillouin zone size.}

\bibitem[{\citenamefont{Mariani and von Oppen}(2008)}]{Oppen:Flexural}
\bibinfo{author}{\bibfnamefont{E.}~\bibnamefont{Mariani}} \bibnamefont{and}
  \bibinfo{author}{\bibfnamefont{F.}~\bibnamefont{von Oppen}},
  \bibinfo{journal}{Phys. Rev. Lett.} \textbf{\bibinfo{volume}{100}},
  \bibinfo{pages}{076801} (\bibinfo{year}{2008}).

\bibitem[{\citenamefont{Mariani and von Oppen}(2010)}]{Oppen:Suspended}
\bibinfo{author}{\bibfnamefont{E.}~\bibnamefont{Mariani}} \bibnamefont{and}
  \bibinfo{author}{\bibfnamefont{F.}~\bibnamefont{von Oppen}},
  \bibinfo{journal}{Phys. Rev. B} \textbf{\bibinfo{volume}{82}},
  \bibinfo{pages}{195403} (\bibinfo{year}{2010}).

\bibitem[{\citenamefont{Ortix et~al.}(2011)\citenamefont{Ortix, Yang, and {van
  den Brink}}}]{Brink:Incommensurate}
\bibinfo{author}{\bibfnamefont{C.}~\bibnamefont{Ortix}},
  \bibinfo{author}{\bibfnamefont{L.}~\bibnamefont{Yang}}, \bibnamefont{and}
  \bibinfo{author}{\bibfnamefont{J.}~\bibnamefont{{van den Brink}}},
  \bibinfo{journal}{arXiv:1111.0399v1}  (\bibinfo{year}{2011}).

\bibitem[{\citenamefont{Park et~al.}(2008)\citenamefont{Park, Yang, Son, Cohen,
  and Louie}}]{Louie:New}
\bibinfo{author}{\bibfnamefont{C.-H.} \bibnamefont{Park}},
  \bibinfo{author}{\bibfnamefont{L.}~\bibnamefont{Yang}},
  \bibinfo{author}{\bibfnamefont{Y.-W.} \bibnamefont{Son}},
  \bibinfo{author}{\bibfnamefont{M.~L.} \bibnamefont{Cohen}}, \bibnamefont{and}
  \bibinfo{author}{\bibfnamefont{S.~G.} \bibnamefont{Louie}},
  \bibinfo{journal}{Phys. Rev. Lett.} \textbf{\bibinfo{volume}{101}},
  \bibinfo{pages}{126804} (\bibinfo{year}{2008}).

\bibitem[{\citenamefont{Kn{\"a}bchen}(1997)}]{Knabchen:SurfaceAcoustic}
\bibinfo{author}{\bibfnamefont{A.}~\bibnamefont{Kn{\"a}bchen}},
  \bibinfo{journal}{Phys. Rev. B} \textbf{\bibinfo{volume}{55}},
  \bibinfo{pages}{6701} (\bibinfo{year}{1997}).

\end{thebibliography}
\end{document}